\documentclass[11pt,a4paper]{article}
\usepackage{jheppub}
\usepackage{mathptmx}
\usepackage{array} 
\usepackage{color}
\usepackage[scaled=0.9]{helvet}
\usepackage{amssymb}

\def\mev{\,\mathrm{MeV}}
\def\gev{\,\mathrm{GeV}}
\def\vec#1{\mathbf{#1}}
\def\fKpi{f_+^{K\pi}}

\newcommand{\fpzero}{\fKpi(0)}

\title{The kaon semileptonic form factor with near physical domain 
	wall quarks}

\collaboration{RBC/UKQCD Collaboration}

\author[a]{P.A.~Boyle}
\author[b]{J.M.~Flynn}
\author[c]{N.~Garron}
\author[b]{A.~J\"uttner}
\author[b]{C.T.~Sachrajda}
\author[a]{K.~Sivalingam}
\author[d]{J.M.~Zanotti}

\affiliation[a]{School of Physics \& Astronomy, University of
  Edinburgh, EH9 3JZ, UK}
\affiliation[b]{School of Physics \& Astronomy, University of
  Southampton, SO17 1BJ, UK}
\affiliation[c]{School of Mathematics, Trinity College, Dublin 2, Ireland}
\affiliation[d]{CSSM, School of Chemistry and Physics, 
	University of Adelaide, SA 5005, Australia}

\emailAdd{juettner@soton.ac.uk}

\abstract{We present a new calculation of the 
  $K\to\pi$ semileptonic form factor at zero
  momentum transfer in domain wall lattice QCD with $N_f =
  2{+}1$ dynamical quark flavours. 
 By using partially twisted boundary
  conditions we simulate directly at the phenomenologically relevant point
  of zero momentum transfer.
We perform a joint analysis for all available ensembles which include three different lattice spacings ($a$\,=\,0.09 -- 0.14\,fm), large physical volumes ($m_\pi L>3.9$) and pion masses as low as 171 MeV.
  The comprehensive set of simulation points allows for a detailed study 
  of systematic effects leading to the prediction $\fpzero=0.9670(20)(^{+18}_{-46})$,
  where the first error is statistical and the second error systematic.
  The result allows us to extract the CKM-matrix element $|V_{us}|=0.2237(^{+13}_{-\;\;8})$
  and confirm first-row CKM-unitarity in the Standard Model 
  at the sub per mille level.
  }


\begin{document}
\maketitle

\section{Introduction}
In the Standard Model (SM), the unitary Cabibbo-Kobayashi-Maskawa (CKM) matrix parametrises the relative strength of different flavour-changing weak processes.
Inconsistencies in the CKM-picture would indicate the presence of new 
physics beyond the SM. 
 It is therefore important to determine all CKM-matrix elements as precisely as
possible by studying flavour changing processes both experimentally 
(e.g. at the NA62 and LHCb experiments at CERN) and theoretically.

In this paper we discuss the determination of the matrix element $|V_{us}|$ from
the study of semileptonic kaon ($K_{l3}$) decays and the test of the unitarity of the first row of the CKM matrix $|V_{ud}|^2 + |V_{us}|^2 + |V_{ub}|^2 = 1$.
$|V_{ub}|$ is very small, O$(10^{-3})$, compared to  the current uncertainties
in $|V_{us}|$ and $|V_{ud}|$. 
The matrix element 
$|V_{ud}|$ is known very precisely from neutron 
$\beta$-decay~\cite{Hardy:2008gy} and
reaching comparable precision for $|V_{us}|$ 
is crucial in searching for deviations
from CKM-unitarity and for possible signs of new
physics.
This can be achieved by combining lattice results for the $K\to\pi$ form
factor $\fpzero$
with $|\fKpi(0)V_{us}|$ from the phenomenological 
analysis~\cite{Antonelli:2010yf} of experimental results. Note that 
$|V_{us}|$ can also be determined from the experimental 
measurement of pion and kaon leptonic decays and lattice results
for the ratio of decay constants $f_K/f_\pi$ 
(cf. FLAG~\cite{Colangelo:2010et}).

The field-theoretical and technical tools developed in the series of papers
~\cite{Hashimoto:1999yp,Becirevic:2004ya,Boyle:2007wg,Boyle:2010bh} have
enabled the calculation of $\fKpi(0)$ in lattice computations
with a precision of around
0.5\%~\cite{Lubicz:2009ht,Boyle:2007qe,Boyle:2010bh,Kaneko:2011rp,Bazavov:2012cd}. The current
experimental uncertainty in $|\fKpi(0) V_{us}|=0.2163(5)$ is about
0.2\%, with an anticipated further reduction of about $30\%$ in this uncertainty
from the KLOE-2-experiment~\cite{AmelinoCamelia:2010me}.
These results challenge lattice simulations to achieve a similar precision.

In our previous work~\cite{Boyle:2007qe,Boyle:2010bh} we have shown
that a precision of 0.5\% can indeed be achieved in practice. We  have also
removed one of the dominant sources of systematic error by using the
method of partially-twisted boundary conditions (explained below) to
avoid an interpolation in the momentum transfer $q^2$ to the point
$q^2=0$~\cite{Boyle:2007wg,Boyle:2010bh}. Other collaborations have computed
the form factor in lattice QCD with $N_f=2$~\cite{Tsutsui:2005cj,Dawson:2006qc,Lubicz:2009ht,Lubicz:2010bv} 
and $N_f=2+1$~\cite{Kaneko:2011rp,Bazavov:2012cd} 
dynamical quarks and an overview of the world data can be found in
the FLAG report~\cite{Colangelo:2010et}. The distinct features of our new
calculations are results for three values of the lattice spacing (a\,=\,0.09\,fm\,--\,0.14\,fm),
lighter simulated quark masses 
($m_\pi=171$MeV)~\cite{Arthur:2012opa}
than used in previous calculations and 
simulations in large volume ($m_\pi L > 3.9$) 
using partially twisted boundary conditions.

In the remainder of the paper we present the details of the calculation, but here we anticipate the final result. The comprehensive set of simulation points allows for a detailed study 
of systematic effects leading to the result:
\begin{equation}
	\fpzero=0.9670(20)(^{+18}_{-46})\,,\qquad
	|V_{us}|=0.2237(^{+13}_{-\;\;8})\,,
\end{equation}
where in the result for the form factor 
the first error is statistical and the second error systematic.
Our result for the CKM matrix element $|V_{us}|$ allows for the 
confirmation of first-row CKM-unitarity in the Standard Model 
at the sub per mille level.

In the following we start with a discussion of the techniques used to determine
the form factor in terms of Euclidean correlation functions. We then explain
our choice of simulation parameters, followed by a description of the calculation itself,  
the extrapolation of the lattice data to the physical point and the
error budget for final results. Finally we present our conclusions.
\section{Calculational procedure}

The matrix element of the vector current between initial and final
pseudoscalar states $P_i$ and $P_f$ decomposes into two form factors,
\begin{equation}
\label{eq:formfactors}
\langle P_f(p_f) | V_\mu | P_i(p_i) \rangle =
 f_+^{P_iP_f}(q^2) (p_i+p_f)_\mu + f_-^{P_iP_f}(q^2) (p_i-p_f)_\mu,
\end{equation}
where $q=p_f-p_i$ is the momentum transfer. For $K\to\pi$ semileptonic
decay, $P_i=K$, $P_f=\pi$ and $V_\mu = \bar s \gamma_\mu u$. The
scalar form factor is defined by
\begin{equation}
f_0^{K\pi}(q^2) = \fKpi(q^2) + \frac{q^2}{m_K^2-m_\pi^2} f_-^{K\pi}(q^2)
\end{equation}
and satisfies $f_0^{K\pi}(0) = \fKpi(0)$.

\begin{figure}
	\begin{center}
	\includegraphics[width=8cm]{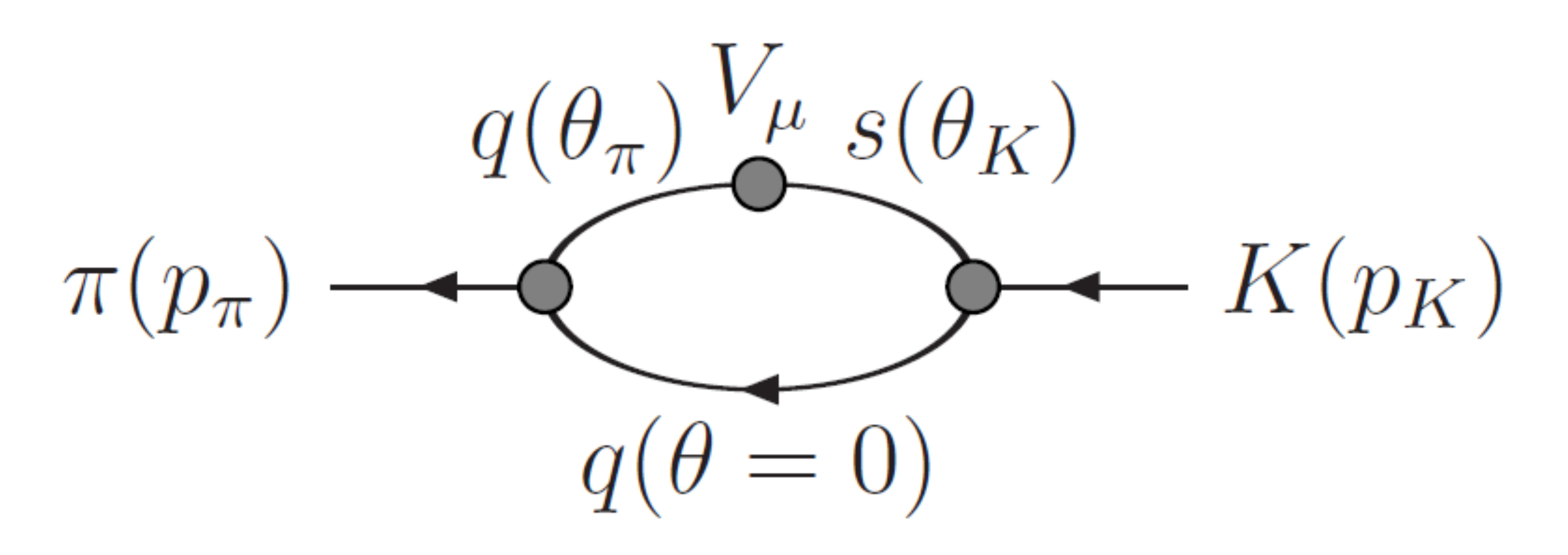}
\end{center}
\caption{Quark flow diagram for a 3pt function with initial and final states
$P_i$ and $P_f$, respectively.}
\label{fig:quarkflow}
\end{figure}

In order to simulate directly at $q^2=0$, we use partially twisted
boundary conditions~\cite{Sachrajda:2004mi,Bedaque:2004ax}, combining
gauge field configurations generated using sea quarks obeying periodic
spatial boundary conditions with valence quarks obeying twisted boundary
conditions. Specifically, the valence quarks satisfy boundary conditions of the form:
\begin{equation}
\psi(x_k + L) = e^{i\theta_k}\psi(x_k)  \qquad k=1,2,3,
\end{equation}
where $\psi$ is either a strange quark or one of the degenerate up and
down quarks. The dispersion relation for a meson in cubic volume
$V=L^3$ projected onto Fourier momentum $\vec{p}_\mathrm{F}$ takes the
form~\cite{deDivitiis:2004kq,Flynn:2005in},
\begin{equation}
E^2 = m^2 + (\vec{p}_\mathrm{F}+\Delta\vec{\theta}/L)^2\,,
\label{eq:pi_disprel}
\end{equation}
where $E$ is the energy, $m$ is the meson's mass and
$\Delta\vec{\theta}$ is the difference of the twist angles for the two
valence quarks in the meson. By varying the twist angles, arbitrary
momenta can be reached. Here we choose the angles such that 
$q^2=0$. In the quark flow diagram of Figure~\ref{fig:quarkflow}, we
twist the strange ($s$) and light quarks ($q$) coupling to the vector current
with phases $\theta_K$ and $\theta_\pi$
in order to give momenta to the kaon and pion respectively. 
The choice of twisting angles is
discussed further in Section~\ref{sec:simparams} and
their values are given in Table~\ref{tab:twists}. 

The matrix element in~\eqref{eq:formfactors} can be extracted from
the time dependence of combinations of Euclidean two- and three-point
correlation functions in lattice QCD. The two-point function is
defined by
\begin{equation}
C_i(t,\vec p_i) \equiv \sum_{\vec{x}}e^{i\vec{p}_i\cdot\vec{x}} \langle
     \,O_i(t,\vec{x})\, O_i^\dagger(0,\vec{0})\,\rangle
   = \frac{|Z_i|^2}{2E_i} \left(e^{-E_it}+ e^{-E_i(T-t)}\right)\, ,
\label{eq:twopt}
\end{equation}
where $i=\pi$ or $K$, and $O_i$ are pseudoscalar interpolating
operators for the corresponding mesons, $O_\pi= \bar q \gamma_5 q$ and
$O_K=\bar s\gamma_5 q$. We assume that $t$ and $T-t$ (where $T$ is the
temporal extent of the lattice) are large enough that the correlation
function is dominated by the lightest state (i.e.\ the pion or kaon).
The constants $Z_i$ are given by
$Z_i=\langle\,P_i\,|\,O_i^\dagger(0,\vec{0})\,|\,0\,\rangle$. The
three-point functions are defined by
\begin{equation}
\begin{aligned}
C^{(\mu)}_{P_iP_f}(t_{i},t,t_{f},\vec p_i,\vec p_f)
 &\equiv Z_V\sum_{\vec{x}_f,\vec{x}} e^{i\vec{p}_f\cdot(\vec{x}_f-\vec{x})}
    e^{i\vec{p}_i\cdot\vec{x}} \langle\, O_f(t_{f},\vec x_f)\,
    V_\mu(t,\vec{x})\,O_i^\dagger(t_{i},\vec 0)\,\rangle\\
 &= Z_V \frac{Z_i\,Z_f}{4E_iE_f}\, \langle\,P_f(\vec{p}_f)\,|\,V_\mu(0)\,|\,
        P_i(\vec{p}_i)\,\rangle\\
 &\phantom{=} \times\left\{\theta(t_f-t)\,e^{-E_i(t-t_i)-E_f(t_f-t)}\ \right.
   +\,c_\mu \left.\theta(t-t_f)\,e^{-E_i(T+t_i-t)-E_f(t-t_f)}\right\}\,,
\end{aligned}
\label{eq:3pt}
\end{equation}
where $P_{i,f}$ is a pion or a kaon, $V_\mu$ is the vector current
with flavour quantum numbers to allow the $P_i\to P_f$ transition and
we have defined $Z_f=\langle\, 0\,|\,O_f(0,\vec 0)|\,P_f\,\rangle$.
The constant $c_\mu$ satisfies $c_0=-1$ (time-direction) and $c_i=+1$
for $i=1,2,3$. Again we assume that all time intervals are
sufficiently large for the lightest hadrons to give the dominant
contribution.

We obtain the vector current renormalisation factor $Z_V$ as follows.
For illustration, take $0< t< t_f <T/2$, in which case $Z_V$ is
defined by
\begin{equation}
\label{eq:zv}
Z_V = \frac{\tilde C_\pi(t_f,\vec 0)}
    {C^{(B,0)}_{\pi\pi}(t_i,t,t_f,\vec{0},\vec{0}\,) }\,.
\end{equation}
In the numerator we use the function $\tilde C_\pi(t,\vec p)=
C_\pi(t,\vec p)- \frac 12 C_\pi(T/2,\vec p)\,e^{-E_\pi (T/2-t)}$ where
$Z_\pi$ and $E_{\pi}$ are determined by fitting $C_\pi(t,\vec 0)$
and applying~\eqref{eq:pi_disprel}. 
The superscript $B$ in the denominator indicates
that we take the bare (unrenormalised) current in the three-point
function.

In the following we drop the labels $t_i$ and $t_f$ in the three-point functions (since they are fixed)
and we combine the two- and three- point functions into the ratios
\begin{equation}
\label{eq:ratios}
\begin{aligned}
R^{(\mu)}_{1}(\vec{p}_K,\vec{p}_\pi)
    &= 4\sqrt{E_K E_\pi}\, \sqrt{\frac
     {C^{(\mu)}_{K \pi}(t,\vec p_K,\vec p_\pi)\,
      C^{(\mu)}_{\pi K}(t,\vec p_\pi,\vec p_K)}
     {\tilde C_{K}(t_\pi,\vec p_K)\,\tilde C_{\pi}(t_\pi,\vec p_\pi)}},
 \\
 R^{(\mu)}_{2}(\vec{p}_K ,\vec{p}_\pi)&=
 2\sqrt{{E_K  E_\pi }}\,
 \sqrt{
 \frac{C^{(\mu)}_{K \pi }(t,\vec p_K ,\vec p_\pi)
 \,C^{(\mu)}_{\pi K }(t,\vec p_\pi ,\vec p_K)}
 {C^{(0)}_{K K }(t,\vec p_K ,\vec p_K)\,
 C^{(0)}_{\pi \pi }(t,\vec p_\pi ,\vec p_\pi)}}\,,\\
\end{aligned}
\end{equation}
which are constructed such that
\begin{equation}
R^{(\mu)}_{\alpha}(\vec{p}_K ,\vec{p}_\pi ) =
  f^{K \pi }_+(q^2)({p}_K +{p}_\pi )_\mu+f^{K \pi }_-(q^2)({p}_K -{p}_\pi)_\mu\,,
\end{equation}
for $\alpha=1,2$. For the ratios we use the naming convention of
\cite{Boyle:2007wg}.

Once these ratios have been computed for several choices for $\vec p_K$
and $\vec p_\pi$ while keeping $q^2$ constant at zero
the form factor $\fpzero$ 
can be obtained as the solution of the corresponding system of linear
equations.

\section{Simulation parameters}
\label{sec:simparams}

We use ensembles of gauge fields with $N_f=2+1$ dynamical flavours at
three different lattice spacings, $a$. The basic parameters of these
ensembles are listed in Table~\ref{tab:ensemble-params}.
On the finer lattices (smaller lattice spacings) we use the
Iwasaki gauge action~\cite{Iwasaki:1984cj,Iwasaki:1985we} and the
domain wall fermion (DWF) action~\cite{Kaplan:1992bt,Shamir:1993zy}.
Two of the ensembles are generated with 
$\beta=2.25$ labelled C in the following, 
$a^{-1}=2.31(4)\gev$)~\cite{Aoki:2010dy} and $\beta=2.13$
labelled A in the following, $a^{-1}=1.75(3)\gev$~\cite{Allton:2008pn}, with
lattice sizes $32^3\times64$ and $24^3\times64$ respectively, with 
the extent of the 
fifth dimension $L_s=16$ in both cases. On these
ensembles we have simulated with unitary pion masses down to 
$295\mev$~\cite{Allton:2008pn,Aoki:2010dy}. To
reach lower unitary pion masses, as low as $171\mev$~\cite{Arthur:2012opa}, 
we have used an additional 
third set of ensembles that employs the Iwasaki gauge action with an
additional weighting factor, the `dislocation suppressing determinant
ratio' (DSDR) in the path integral to suppress gauge configurations
with localised instanton-like artifacts. The dislocations support
additional low-modes of the Dirac operator and suppressing them helps
reduce the growth of the residual mass $m_\mathrm{res}$ (which
parameterises the explicit chiral symmetry breaking arising from the
finite fifth dimension) on this coarser lattice. The Iwasaki-DSDR
ensembles are generated at $\beta=1.75$ labelled B, $a^{-1}=1.37(1)$GeV, 
with lattice size
$32^3\times64$ and $L_s=32$. On all ensembles the spatial volumes are
large enough to ensure $m_\pi L\,{\gtrsim}\, 3.9$ for all simulated masses,
keeping finite-volume corrections small.  
More details on all these ensembles and results for a variety of light
hadronic quantities are given
in~\cite{Allton:2007hx,Allton:2008pn,Aoki:2010dy,Aoki:2010pe,Arthur:2012opa}.
We note that owing to the change in action, cutoff-effects which start at
$O(a^2)$ for DWF will behave differently for ensemble B than for 
A and C~\cite{Arthur:2012opa}. 
This will be discussed further in  Section~\ref{sec:extrapolations}.

\begin{table}
\begin{center}
        \small
\begin{tabular}{l@{\hspace{3mm}}l@{\hspace{2mm}}c@{\hspace{2mm}}c@{\hspace{2mm}}c@{\hspace{2mm}}c@{\hspace{2mm}}ccccc@{\hspace{2mm}}c@{\hspace{2mm}}c@{\hspace{2mm}}c}
 \hline\hline &\multicolumn{2}{c}{action}&\\[-0.4ex]
 set & sea & val & $\beta$ & $a\,/\mathrm{fm}$ & $L/a$ & $T/a$
 & $am_q$ & $am_s^\mathrm{sea}$ & $am_s^{\rm val}$&$m_\pi$/MeV&$N_{\rm src}$&$N_{\rm meas}$&$m_\pi L$\\ \hline\\[-4mm]
 A$_3$   &DWF &DWF&2.13 &0.11 &24 &64 &0.0300&0.040&0.040 &678&2 &105&9.3\\
 A$_2$   &DWF &DWF&2.13 &0.11 &24 &64 &0.0200&0.040&0.040 &563&2 &85 &7.7\\
 A$_1$   &DWF &DWF&2.13 &0.11 &24 &64 &0.0100&0.040&0.040 &422&2 &153&5.8\\
 A$_5^4$ &DWF &DWF&2.13 &0.11 &24 &64 &0.0050&0.040&0.040 &334&8 &143&4.6\\
 A$_5^3$ &DWF &DWF&2.13 &0.11 &24 &64 &0.0050&0.040&0.030 &334&8 &143&4.6\\
 C$_8$   &DWF &DWF&2.25 &0.09 &32 &64 &0.0080&0.030&0.025 &398&8 &120&5.5\\
 C$_6$   &DWF &DWF&2.25 &0.09 &32 &64 &0.0060&0.030&0.025 &349&8 &153&4.8\\
 C$_4$   &DWF &DWF&2.25 &0.09 &32 &64 &0.0040&0.030&0.025 &295&9 &135&4.1\\
 B$_{42}$   &DSDR&DWF&1.75 &0.14 &32 &64 &0.0042&0.045&0.045 &248&16 &162&5.7\\
 B$_1$   &DSDR&DWF&1.75 &0.14 &32 &64 &0.0010&0.045&0.045 &171&16 &196&3.9\\
 \hline\hline
\end{tabular}
\end{center}
\caption{Basic parameters for all ensembles of gauge field configurations.}
\label{tab:ensemble-params}
\end{table}

For the computation of the form factor we  distinguish two different 
kinematical situations.
We denote by `kinematics I' the case where 
either the kaon or the pion are at rest~\cite{Boyle:2007wg}:
\begin{align}
|\vec{\theta}_K| &=
 L \sqrt{\left(\frac{m_K^2+m_\pi^2}{2m_\pi}\right)^2 -m_K^2}
 \qquad\text{and}\qquad \vec{\theta}_\pi=\vec{0}\,, \\
|\vec{\theta}_\pi| &=
L \sqrt{\left(\frac{m_K^2+m_\pi^2}{2m_K}\right)^2 -m_\pi^2}
 \qquad\text{and}\qquad \vec{\theta}_K=\vec{0}\,.
\end{align}
In some cases, we twist in more than
one direction. This increases the number of equations from
which we can determine the form factors in~\eqref{eq:formfactors}. 
It may also reduce discretisation errors
coming from $(pa)^2$ and higher powers, where $p$ is a component of the momentum. On the coarsest lattice, twisting only the kaon leads to
large twisting angles, giving the kaon a momentum of order of a typical Fourier
momentum unit, $(2\pi/L)$. Therefore, in `kinematics II' we choose two components of 
the kaon twisting angle to be non-zero but not too large
and then fix the twist of
the pion in the remaining direction to ensure that $q^2=0$.
Our complete set of twisting angles is summarised in 
Table~\ref{tab:twists}.

\begin{table}
\begin{center}
\newcolumntype{R}{@{,\hspace{0.2em}}r}
\begin{tabular}{lrRRr@{\hspace{0mm}}RRrRRrRRrrrr}
\hline\hline
 & \multicolumn6c{kinematics I} & \multicolumn6c{kinematics II}\\
set & \multicolumn3c{$\vec\theta_\pi$}
    & \multicolumn3c{$\vec\theta_K$}
    & \multicolumn3c{$\vec\theta_\pi$}
    & \multicolumn3c{$\vec\theta_K$}\\
\hline
 A$_3  $ & (0.375 & 0.375& 0.375) &(0.402 & 0.402 & 0.402)&\multicolumn{3}{c}{na.}&\multicolumn{3}{c}{na.}     \\
 A$_2  $ & (0.790 & 0.790& 0.790) &(0.943 & 0.943 & 0.943)&\multicolumn{3}{c}{na.}&\multicolumn{3}{c}{na.}     \\
 A$_1  $ & (1.270 & 1.270& 1.270) &(1.842 & 1.842 & 1.842)&\multicolumn{3}{c}{na.}&\multicolumn{3}{c}{na.}     \\
 A$_5^4$ & (2.682 & 0.000& 0.000) &(4.681 & 0.000 & 0.000)&\multicolumn{3}{c}{na.}&\multicolumn{3}{c}{na.}     \\
 A$_5^3$ & (2.129 & 0.000& 0.000) &(3.337 & 0.000 & 0.000)&\multicolumn{3}{c}{na.}&\multicolumn{3}{c}{na.}     \\
 C$_8$   & (0.943 & 1.622& 0.000) &(0.000 & 1.570 & 2.094)&\multicolumn{3}{c}{na.}&\multicolumn{3}{c}{na.} \\
 C$_6$   & (0.943 & 1.934& 0.000) &(0.000 & 1.570 & 2.915)&\multicolumn{3}{c}{na.}&\multicolumn{3}{c}{na.} \\
 C$_4$   & (1.739 & 1.739& 0.000) &(0.000 & 3.086 & 3.086)&\multicolumn{3}{c}{na.}&\multicolumn{3}{c}{na.} \\
 B$_{42}$   & (3.209 & 0.000& 3.209) &(0.000 & 6.587 & 6.587)

         & (3.689 & 0.000& 0.000) &(0.000 & 2.356 & 3.927)         \\
 B$_1$   & (2.513 & 4.382& 0.000) &(0.000 & 0.000 & 0.000)
         & (0.000 & 0.000& 4.173) &(4.712 & 3.142 & 0.000)         \\

\hline\hline
\end{tabular}
\end{center}
\caption{Choice of twist angles $\vec \theta$ for kinematical configurations I
and II (na. indicates the cases where data for only one type of kinematics was
generated).}
\label{tab:twists}
\end{table}
\begin{table}
\begin{center}
\begin{tabular}{lllllllll}
 \hline\hline
 set 	&$am_\pi$ &$am_K$ & $\fpzero$\\[-0mm]
 \hline
 A$_3$ &0.38838(39)&0.41626(39)& 0.9992(1)\\
 A$_2$ &0.32234(47)&0.38437(48)& 0.9956(4)\\
 A$_1$ &0.24157(40)&0.35009(41)& 0.9870(9) \\
 A$_5^4$ &0.19093(45)&0.33198(58)& 0.9760(43) \\
 A$_5^3$ &0.19093(45)&0.29819(52)& 0.9858(28) \\
 C$_8$   &0.17247(49)&0.24123(47)& 0.9904(17) \\
 C$_6$   &0.15105(44)&0.23274(47)& 0.9845(23) \\
 C$_4$   &0.12776(43)&0.22623(54)& 0.9826(35) \\
 B$_{42}$&0.18067(19)&0.37157(29)& 0.9771(21) \\
 B$_1$   &0.12455(20)&0.35920(31)& 0.9710(45) \\
 \hline\hline
\end{tabular}
\end{center}
\caption{Measured properties on all ensembles}\label{tab:ffresults}
\end{table}

\section{Numerical results}\label{sec:numerical results}
We have used the bootstrap procedure~\cite{efron:1979} with 500 bootstrap
samples for each ensemble. The choice of binning is based on previous
auto-correlation studies in~\cite{Allton:2008pn,Aoki:2010dy,Arthur:2012opa}.
We have combined the measurements for $N_{\rm src}$ noise source positions 
(cf. Table~\ref{tab:ensemble-params})
into one bin, in this way achieving a sufficient {effective} 
separation between subsequent configurations in molecular dynamics time.
All two-point and three-point correlation functions were computed using the
stochastic source technique~\cite{Foster:1998vw,McNeile:2006bz,Boyle:2008rh}
with one \textit{hit} per source position.

We determine the pion and kaon masses for each ensemble from fits to the two-point correlation function \eqref{eq:twopt}; the results are summarised in 
Table~\ref{tab:ffresults}. The table also contains the results for 
the form factors $\fpzero$. These were determined as the solution
of an over-constrained system of linear equations composed of all results
of constant fits to the ratios $R_1^{(\mu)}$ and 
$R_2^{(\mu)}$ ($\mu=0,1,2,3$) at $q^2=0$ on a given ensemble. 
No further interpolation in the momentum transfer was necessary.
As mentioned in the previous section `kinematics I' leads to rather large
twist angles for the kaon in the case where the pion is at rest and its mass
closer to the physical point. As the momentum is increased 
the signal quality deteriorates to the extent
that in some cases a clear identification
of a plateau region is impossible.
Before solving the system of linear equations we inspected each individual
ratio-fit and discarded those results which were of unsatisfactory 
quality. 

We note that measurements labelled with A$_5^4$ and A$_5^3$ are 
based on the same ensemble of gauge configurations and differ only in the 
choice of the strange quark mass ($am_s=0.04$ is unitary while 
$am_s=0.03$ is partially quenched).
With the exception of these two cases
all results are fully independent, i.e. no statistical correlations are present.
\section{Extrapolations}\label{sec:extrapolations}
Each individual simulation of lattice QCD differs from the strong 
interaction found in nature. In addition to
$SU(2)$-isospin breaking effects, which are beyond the scope of this paper,
this is predominantly because computer simulations 
are naturally limited to finite volumes and lattice spacings. 
Moreover, the simulated quark masses do not 
correspond exactly to the physical ones. Over the years we have gained 
experience 
in dealing with the resulting systematic effects, very often guided  
by predictions of effective field theories.
In this section we discuss the extrapolation of the lattice data
to the physical point corresponding to the
$K^0\to\pi^-$ decay defined in terms of the 
charged pion mass $m_{\pi^-}=139.57$\,MeV and the neutral kaon mass 
$m_{K^0}=497.614$\,MeV~\cite{Beringer:1900zz}.

We start by briefly recalling the prediction for the form factor in 
chiral perturbation theory,
\begin{equation}\label{eq:fpzeroexpansion}
	\fpzero=1+f_2(f,m_\pi^2,m_K^2,m_\eta^2)+\Delta_f\,,
\end{equation}
where from~\cite{Gasser:1984ux},
\begin{equation}\label{eq:f2}
	\begin{array}{c}
	f_2(f,m_\pi^2,m_K^2,m_\eta^2)=\frac 32
		H(f,m_\pi^2,m_K^2)+\frac 32 H(f,m_\eta^2,m_K^2)\,,\\[2mm]
		{\rm with}\,\,	H(f,m_P^2,m_Q^2)=-\frac 1{64\pi^2f^2}
		\left(
		m_P^2+m_Q^2+2\frac{m_P^2 m_Q^2}{m_P^2-m_Q^2}
		\log\left(\frac {m_Q^2}{m_P^2}\right)\right)\,,
\end{array}
\end{equation}
is the next-to-leading order (NLO) contribution which, apart from the 
pseudo-scalar decay constant $f$, is
parameter-free~\cite{Ademollo:1964sr,Gasser:1984ux}. $\Delta_f$ represents 
next-to-next-to-leading order (NNLO) contributions (computed by
Bijnens and Talavera~\cite{Bijnens:2003uy})
and beyond. 
Expression \eqref{eq:fpzeroexpansion} respects the $SU(3)$-symmetry,
$\fpzero=1|_{m_s=m_l}$, and deviations from this limit start proportional to   
$(m_K^2-m_\pi^2)^2/m_K^2$. 
In the following we employ the tree-level relation 
$m_\eta=\sqrt{(4m_K^2-m_\pi^2)/3}$ for the $\eta$-mass.

\begin{figure}
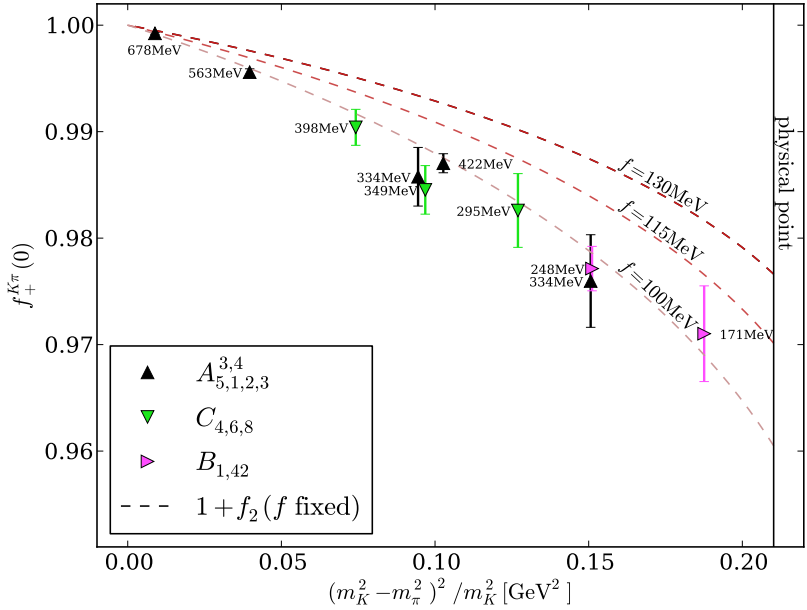

	\begin{center}
	\includegraphics[width=12cm]{{{plots/bare_results}}}
	\end{center}
\caption{Illustration of simulation results for the $K\to\pi$ form factor
	from all ensembles 
	together with the prediction of chiral perturbation theory
	at NLO for different representative choices of the 
	input value for the  decay constant $f$ (dashed lines).
	}\label{fig:bare results}
\end{figure}

The ratios $R_1^{(\mu)}$ and $R_2^{(\mu)}$ in \eqref{eq:ratios} from which 
we compute the form factor are constructed such that
$\fpzero|_{m_s=m_l}=1$ holds exactly even  
in a finite volume and for a finite lattice cut-off. We therefore expect
finite-volume and cut-off effects to be symmetry-suppressed.  Because of the automatic $O(a)$-improvement with domain wall Fermions
on the coarse ensemble (B) we expect  
$O\left((a\Lambda_{\rm QCD})^2\right)\approx 5\%$ 
cut-off effects on the deviation of
the form factor from one and on the finest ensemble (C) we expect these
effects to be around 2\%, where we have assumed 
$\Lambda_{\rm QCD}\approx 300$\,MeV.
A very good numerical confirmation that these effects are indeed
below the statistical precision of our simulations can be 
seen in Figure~\ref{fig:bare results}. The plot  
provides a first impression of the 
simulation results plotted against $(m_K^2-m_\pi^2)^2/m_K^2$ 
($\fpzero$ should be linear in this variable for $m_K^2$ close to $m_\pi^2$).
The data points for $m_\pi=248$\,MeV (B$_{42}$) and $m_\pi=334$\,MeV (A$_5^4$,
see pion mass labels in the plot)
with $a=0.14$\,fm and $a=0.11$\,fm, respectively,  lie on top 
of each other.
Similarly, the $m_\pi=334$\,MeV (A$_5^3$) and $m_\pi=349$\,MeV (C$_6$) 
simulation points
for $a=0.11$\,fm and $a=0.09$\,fm, respectively, are in complete agreement
and cut-off effects are therefore absent at the current level of precision.

The chiral effective theory for the quantities considered here predicts
finite volume effects to be exponentially 
suppressed~\cite{Ghorbani:2011gc,Ghorbani:2013yh} (proportional to $e^{-m_\pi L}$).
The values of $m_\pi L$ are summarised
in Table~\ref{tab:ensemble-params} and they are all larger than 3.9. 
Finite volume effects are therefore expected
to be of order 2\% and below. Again, this uncertainty affects only the  difference of the form 
factor from one which is a tiny effect. In this section
we therefore assume that both lattice artifacts and finite-volume effects are below the statistical accuracy of the results. Further considerations will follow in the next
section.

\begin{figure}
	\begin{center}
		\includegraphics[width=15cm]{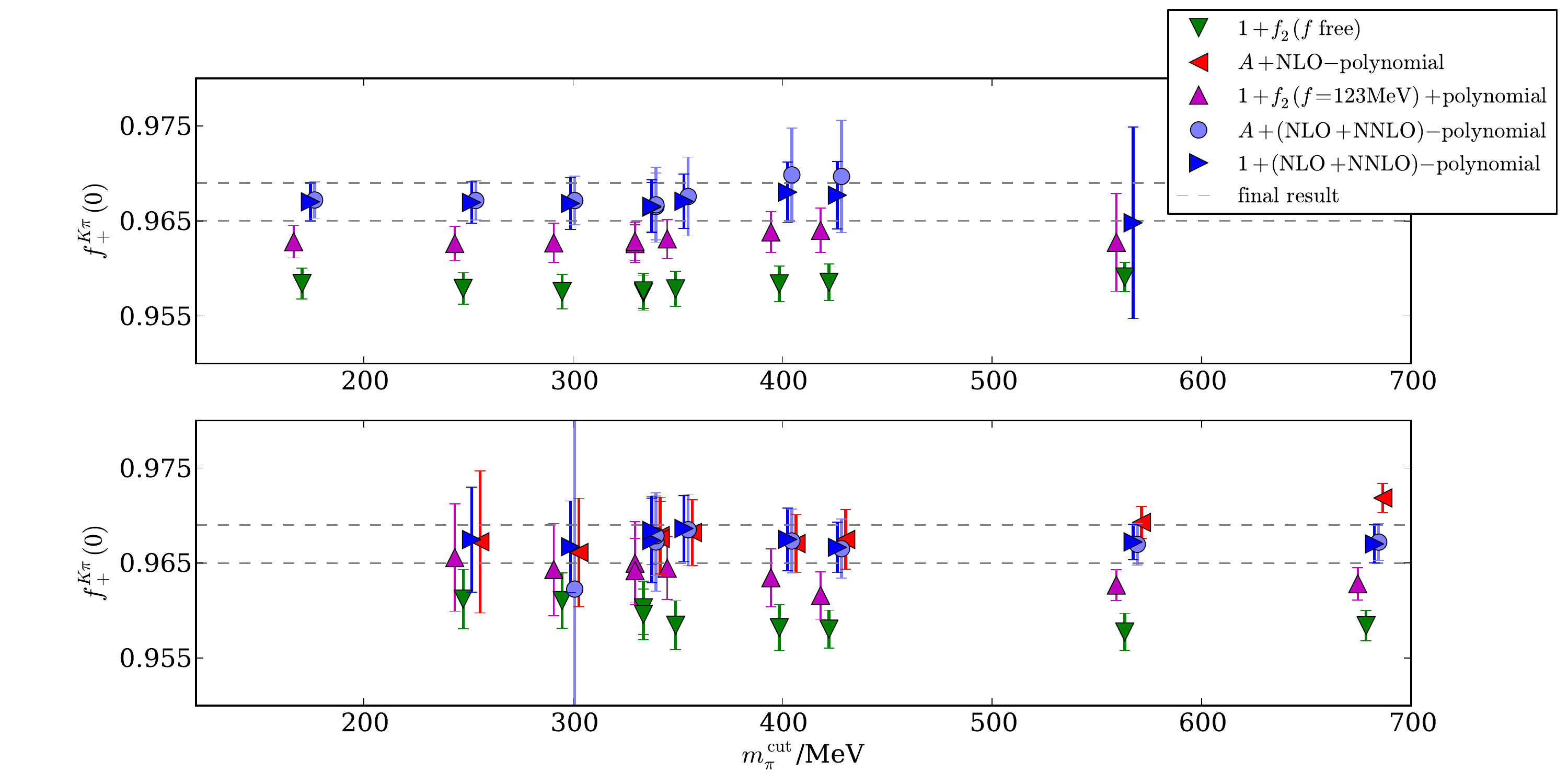}\hspace*{-5mm}
	\end{center}
\caption{Illustration of the dependence of the results for the form
factor on different ans\"atze under the variation of the lightest (top panel) 
and heaviest (bottom plot) 
included pion mass, respectively.}\label{fig:cutoff_dependence}
\end{figure}

\begin{table}
 \begin{center}
  \begin{tabular}{lllcc}
   \hline\hline\\[-4mm]
   id&\multicolumn{2}{l}{fit ansatz for $\fpzero$}&$\fpzero$&$\chi^2/$d.o.f\\
   \hline\\[-4mm]
   $\mathcal{A}$&&$1+f_2(f={\rm free})$			&0.9584(16)&0.6\\
   $\mathcal{B}$&(\ref{eq:f2 plus model}) &$1+f_2(f=95{\rm MeV})+$polynomial	&0.9565(17)&0.7\\
   $\mathcal{C}$&(\ref{eq:f2 plus model})&$1+f_2(f=115{\rm MeV})+$polynomial	&0.9615(17)&0.3\\
   $\mathcal{D}$&(\ref{eq:f2 plus model})&$1+f_2(f=130{\rm MeV})+$polynomial	&0.9639(18)&0.4\\
   $\mathcal{E}$&(\ref{eq:polynomialA})&$A+$polynomial				&0.9672(19)&0.2\\
   $\mathcal{F}$&(\ref{eq:polynomialB})&$1+$polynomial				&0.9670(20)&0.2\\
   \hline\hline\\
  \end{tabular}
 \end{center}
 \caption{Fit results for various fit-functions as discussed in 
 Section~\ref{sec:extrapolations}. In each case the quoted result 
 is for the fit to all simulated data points.}
 \label{tab:results}
\end{table}
\begin{figure}
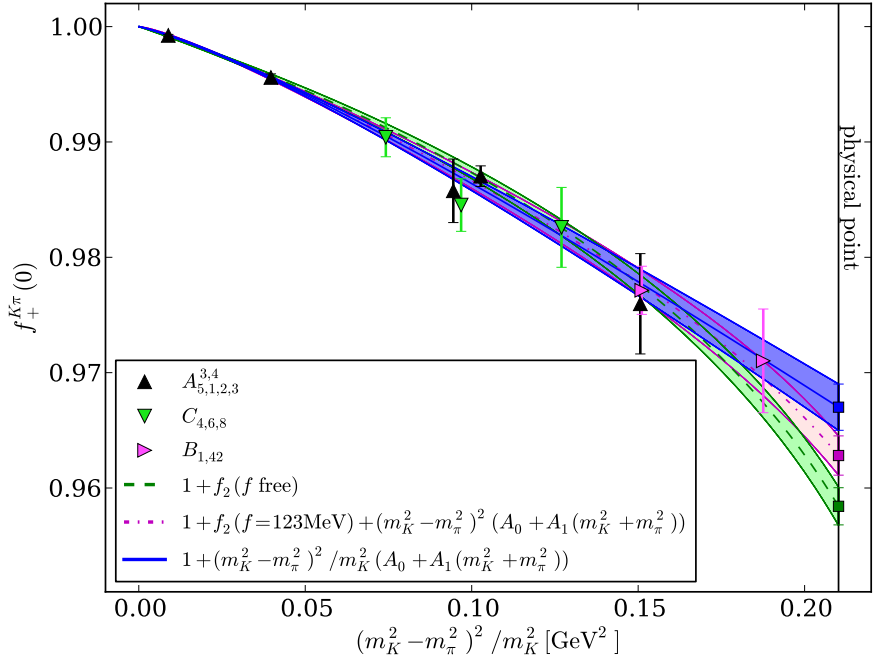

	\begin{center}
	\includegraphics[width=13cm]{{{plots/final_fpi=0.095_9}}}
	\end{center}
\caption{Illustration of results for three different fit-models. In each case
 	the fit to the full set of simulation results is shown.
	}\label{fig:fit plot}
\end{figure}
Figure~\ref{fig:bare results} also shows the prediction of chiral 
perturbation theory at NLO for comparison, 
i.e. the expression $1+f_2(f,m_\pi^2,m_K^2,m_\eta^2)$, which depends
strongly on the choice for the value of the decay constant  used
as input (shown for $f=100,115,130$\,MeV).
As already noted in~\cite{Boyle:2010bh} different choices 
for the value of the decay constant correspond to different forms for the
NNLO-effects. In the
full chiral expansion such effects are compensated by a change in the
decay-constant's contribution at higher order.
Surprisingly, for a value of around $f=100$\,MeV all the results
 seem to be  reasonably
well described by the NLO-ansatz  without any NNLO corrections. We have therefore  
attempted to determine the decay constant from a fit to the data of only
$1+f_2$ (fit $\mathcal{A}$).
The fits were of good quality (cf. the $\chi^2/$d.o.f.-values in 
Table~\ref{tab:results}).
The functional form of fit $\mathcal{A}$ is shown in Figure~\ref{fig:fit plot} 
(dashed central line and green error band) and
Figure~\ref{fig:cutoff_dependence} illustrates how the fit result changes for
different choices of the data points included
(upside-down green triangles). 
The top panel shows how the results depend on variations of the
lowest pion mass included into the fit (while including all heavier data points)
and the bottom plot shows how the
results change as the mass of the heaviest pion included into the fit 
is reduced (while including all results down to the lightest data point).
While the central value of the form factor extrapolated to the physical point 
remained surprisingly stable given the simplicity
of the fit function the results for the decay constant as a fit-parameter 
varied significantly between $f=97$\,MeV and 101\,MeV (not shown).
 As data points closer to the
$SU(3)$-symmetric limit are removed from the fit
the central value  moves a little
towards larger values.  
 
Given the wide range of simulated pion masses the
performance of the ansatz $1+f_2$ is  perhaps accidental
and  higher order terms in the chiral expansion play a role. 
The number of free parameters in the full
NNLO-expression~\cite{Bijnens:2003uy} contained in $\Delta_f$ is, however,
too large to allow for a meaningful fit without further external 
constraints~(cf. MILC~\cite{Bazavov:2012cd} who in their analysis of lattice
data for the form factor constrain NNLO fits by using 
the $O(p^6)$ low-energy constants from Bijnen's ``\textit{Fit 10}'' of 
chiral perturbation theory to experimental data~\cite{Bijnens:2003uy}).
In order to study potential higher order effects 
we employ a model~\cite{Boyle:2007qe,Boyle:2010bh},
\begin{equation}\label{eq:f2 plus model}
          \fpzero=1+f_2(f,m_\pi^2,m_K^2,m_\eta^2) +(m_K^2-m_\pi^2)^2
          \left( A_0+A_1(m_K^2+m_\pi^2) \right)\,.
\end{equation}
 In addition to the dependence on the 
decay constant $f$ this expression has two further parameters, 
$A_0$ and $A_1$. 
Allowing all three parameters to vary did not lead to stable 
fits -- the final
result showed a dependence on the choice of starting values 
for the $\chi^2$-minimisation.
We have therefore tried fits with the decay constant $f$ 
held fixed for a number of 
values in the range 95\,MeV to 130\,MeV (fits $\mathcal{B}$, $\mathcal{C}$ and
$\mathcal{D}$).
For a given choice of $f$ the results for $\fpzero$ 
are of good quality and stable under variation of the 
mass-cuts (cf. Figure~\ref{fig:cutoff_dependence} for $f=$123\,MeV). We did
however find a monotonic variation of $\fpzero$ from 0.9565(17) to $0.9639(18)$ 
when increasing the value of $f$ from  95\,MeV to 130\,MeV (in each case fit to all 
ensembles). We have observed  this behaviour 
previously in~\cite{Boyle:2010bh} where the final
result for $\fpzero$ was determined from the fit of ansatz 
(\ref{eq:f2 plus model}) to data set A with $f=115$\,MeV and a systematic
uncertainty due to the dependence on the decay constant was estimated from
the variation of the central value between $f=100$\,MeV and $f=131$\,MeV.
We note that the result $A_0=-0.32(5)$GeV$^{-4}$ and $A_1=0.26(6)$GeV$^{-6}$
for fit $\mathcal{D}$ agrees with what we found 
for the fit to data set A alone in~\cite{Boyle:2010bh}, 
$A_0=-0.34(9)$GeV$^{-4}$ and $A_1=0.28(12)$GeV$^{-6}$.

As an alternative fit-form a naive
polynomial parameterisation of the data seems equally appropriate, 
particularly as 
the lattice data further away from the 
$SU(3)$-symmetric point shows little or no curvature.
We therefore also consider the fit-ans\"atze ({fit $\mathcal{E}$}
and $\mathcal{F}$),
\begin{eqnarray}
	\fpzero&=&A+\frac{(m_K^2-m_\pi^2)^2}{m_K^2}
          A_0\,,\label{eq:polynomialA}\\
	\fpzero&=&A+\frac{(m_K^2-m_\pi^2)^2}{m_K^2}
          \left( A_0+A_1(m_K^2+m_\pi^2) \right)\,.\label{eq:polynomialB}
\end{eqnarray}
These expressions are motivated by the expansion of \eqref{eq:fpzeroexpansion}
in $(m_K^2-m_\pi^2)$. The parameter $A$ can either be set to 1, i.e. to the
$SU(3)$-symmetric limit or it can be left floating. 
The ansatz~(\ref{eq:polynomialA}) 
leads to acceptable fit-quality only when the heavier data points are 
excluded (otherwise we found large values for $\chi^2/$d.o.f.). 
We have added the corresponding results only in the bottom panel of 
Figure~\ref{fig:cutoff_dependence}.
With the exception of the right-most point $\chi^2/$d.o.f.$<$1, indicating
that the linear ansatz describes the data well.
Ansatz~(\ref{eq:polynomialB}) with $A$ fixed to 1 leads
to mutually compatible and good-quality fits over the full range of mass-cuts. 
Leaving $A$ freely floating has little impact on the fit results. We find
$A=1.00004(24)$ for the fit over all data points which underlines the
compatibility of our data with $SU(3)$-symmetry and the absence of 
cut-off and finite volume effects in this limit to very high precision.

Fit $\mathcal{E}$ and $\mathcal{F}$ are very stable under a change of the
mass-cut (cf. Figure~\ref{fig:cutoff_dependence}, light blue circles and
right-pointing blue triangles, respectively) 
with at the same time very small $\chi^2/$d.o.f..
We find the fit over all ensembles with ansatz~(\ref{eq:polynomialB}) most
convincing.
The preferred fit-ansatz of our earlier study in~\cite{Boyle:2010bh} which
was based on data set A only was~(\ref{eq:f2 plus model}). This ansatz
is also compatible with the data and we use it in the following
for an estimate of the model dependence.
\section{Error budget and final results}
In this section
we estimate the magnitude of systematic uncertainties that need to be
included in a comprehensive error budget.
Following the discussion in Section~\ref{sec:extrapolations} we 
estimate finite volume errors to be of order 2\% and cutoff effects to be of 
order 5\% on the difference of $\fpzero$ from 1. 
The simulation results indicate that these effects are indeed
very small and below the level of statistical uncertainty.
In order to remain conservative we attach a 5\% error on $1-\fpzero$ for
residual cutoff effects. We proceed in the same way 
for finite-size effects for which we attach a 2\% error on $1-\fpzero$.
Uncertainties from the setting of the relative scale of the ensembles are
reflected in the error on the lattice spacing which we have folded into the
bootstrap analysis and are therefore included in the statistical error.

The chiral extrapolation for our final result
is based on fit $\mathcal{F}$ but ansatz~(\ref{eq:f2 plus model})
appears to be  an adequate alternative with the caveat of the dependence on the
external input $f$. When varying the input $f$ in the fit with 
(\ref{eq:f2 plus model}) the value of 
$\chi^2/$d.o.f. has a flat minimum around $f=123$\,MeV. For this value of $f$
we find $\fpzero=0.9628(17)$ and we take the difference in 
central value between this fit result and fit $\mathcal{F}$ 
as the residual model-dependence.
After these considerations our final result is,
\begin{equation}\label{eq:final result}
	\begin{array}{rcc@{\hspace{0mm}}c@{\hspace{0mm}}c@{\hspace{0mm}}c@{\hspace{0mm}}c@{\hspace{0mm}}l@{\hspace{0mm}}l@{\hspace{0mm}}l@{\hspace{0mm}}ll}
	\fpzero&=&0.9670&(20)_{\rm stat}
	&(^{+\;\;0}_{-42})_{\rm model}
	 	      &( 7)_{\rm FSE}
	 	      &(17)_{\rm cutoff}\\
	&& 	      &0.2\% 
		      &0.4\%
	 	      &0.07\%
	 	      &0.2\% \\
		      &=&\multicolumn{5}{l}{0.9670(20)(^{+18}_{-46})\,,}
	\end{array}
\end{equation}
where  in the last line we have added all systematic errors in quadrature.
Our previous result~\cite{Boyle:2010bh} was based on data sets A with fit
ansatz~(\ref{eq:f2 plus model}) where we were very cautious about 
the curvature suggested by the $f_2$-term
as one moves away from the $SU(3)$-symmetric limit. 
We varied the value of the decay constant entering $f_2$ in order 
to quantify the induced systematic uncertainty. The result was 
$0.9599(34)(^{+31}_{-43})(14)$. The central value is fully compatible with 
the same fit applied to the enlarged data set, fit $\mathcal{C}$.

The first applications of our result are predicting the
CKM-matrix element $|V_{us}|$ and testing the unitarity
of the CKM-matrix which is a crucial Standard Model test.
In~\cite{Antonelli:2010yf} the experimental data for $K\to\pi$ semileptonic
decays was analysed. Their result $|V_{us}\fpzero|=0.2163(5)$ combined with our result for $\fpzero$ gives
\begin{equation}\label{eq:Vus}
	|V_{us}|=0.2237(^{+13}_{-\;\;8})\,.
\end{equation}
Together with 
the result $|V_{ud}|=0.97425(22)$~\cite{Hardy:2008gy} from super-allowed
nuclear $\beta$-decay and 
$|V_{ub}| = 4.15(49)\cdot 10^{-3}$~\cite{Beringer:1900zz}
we then confirm CKM-unitarity at the sub per mille level,
\begin{equation}
	|V_{ud}|^2+|V_{us}|^2+|V_{ub}|^2-1=-0.0008(^{+7}_{-6})\,.
\end{equation}
\section{Discussion and Summary}
\begin{figure}
	\begin{center}
	\epsfig{scale=.5,file=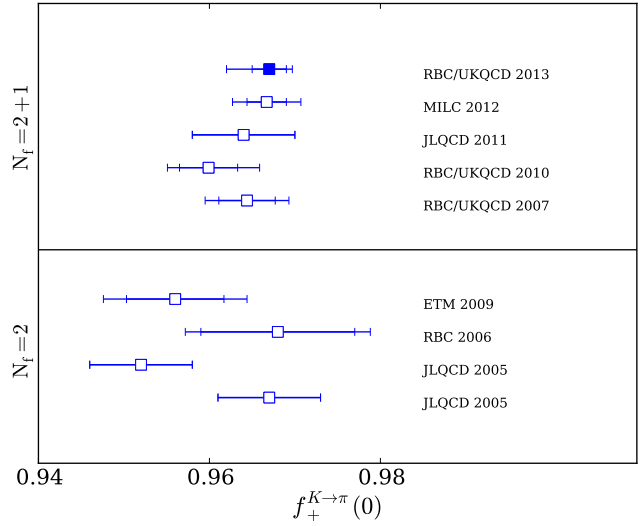}	
	\end{center}
	\caption{Summary of recent lattice results with $N_f=2$ and $N_f=2+1$ 
	dynamical flavours. Where applicable the smaller errorbar corresponds
	to the statistical error only.}\label{fig:worlddata}
\end{figure}
This work constitutes a comprehensive study of the kaon semileptonic decay
form factor in three-flavour lattice QCD. Simulations in large lattice volumes
with three values of the lattice spacing and pion masses in the range
from as low as 171MeV up towards the $SU(3)$-symmetric point allow for 
the detailed study of systematic effects. We have analysed the data using
various ans\"atze for the remaining extrapolation to the physical point 
and we have identified a preferred functional form. 
After the extrapolation to the physical point we
obtain the form factor with a statistical precision of 2 per mille 
and estimated $^{+2}_{-5}$ per mille systematic errors.
The prediction for the form factor, $\fpzero=0.9670(20)(^{+18}_{-46})$ 
has an overall
uncertainty of $^{+0.3}_{-0.5}\%$, where statistical and systematic uncertainties have
been added in quadrature. Our collaboration is currently working on 
supplementing the data set by simulations performed directly at the physical point.
These additional data will allow  us to reduce the dominant
systematic uncertainty, that due to the extrapolation in the quark mass to the 
physical point, very significantly. 
An overview of recent lattice results for the $K\to\pi$ form factor
including our new result is given in Figure~\ref{fig:worlddata}.

An immediate phenomenological application of our result is the test of 
first-row CKM-matrix unitarity in the Standard Model which we are able to
confirm at the sub per mille level. 

\acknowledgments
We are grateful to our colleagues within the RBC/UKQCD collaboration
for sharing resources and gauge ensembles. 
The authors gratefully acknowledge computing time granted 
through the STFC funded DiRAC 
(Distributed Research utilising Advanced Computing) facility 
(grants ST/K005790/1, ST/K005804/1, ST/K000411/1, ST/H008845/1), 
through the Gauss Centre for Supercomputing (GCS) 
	(via John von Neumann Institute for Computing (NIC)) on the 
	GCS share of the supercomputers JUGENE/JUQUEEN at J\"ulich 
	Supercomputing Centre (JSC) and the 
	Engineering and Physical Sciences Research Council (EPSRC) for 
	substantial allocation of time on HECToR under the Early User 
	initiative
and on the QCDOC supercomputer at the University of Edinburgh.
AJ acknowledges funding from the European
Research Council under the European Community's Seventh Framework Programme
(FP7/2007-2013)  ERC grant agreement No 279757.
JMF and CTS acknowledge support from STFC Grant ST/G000557/1 and 
EU contract MRTN-CT-2006-035482 (Flavianet). 
PAB and NG acknowledge support from STFC Grant ST/J000329/1 and KS
is supported by
the European Union under the Grant Agreement number 238353 (ITN STRONGnet).
JMZ is supported by the Australian Research Council grant FT100100005. 

\bibliographystyle{JHEP}
\bibliography{kl3-2012}

\end{document}